\documentclass[unsortedaddress]{revtex4-1}

\usepackage{amsmath}
\usepackage{amsfonts}
\usepackage{amsthm}
\usepackage{amssymb}
\usepackage{graphicx}
\usepackage{hyperref}

	\newcommand{\ncd}{\newcommand}
	\ncd{\mrm}    {\mathrm}
	\ncd{\beq} {\begin{equation}}
	\ncd{\eeq} {\end{equation}}

	\def\d{{\rm d}}

\begin{document}

	\title{Two-dimensional Einstein manifolds in geometrothermodynamics}

	\author{A. C. Guti\'errez-Pi\~{n}eres}
		\email{acgutierrez@correo.nucleares.unam.mx}
		\affiliation{Instituto de Ciencias Nucleares, Universidad Nacional Autonoma de M\'exico,\\
A.P. 70-543, 04510 Mexico D.F., Mexico}
\affiliation{Facultad de Ciencias B\'asicas, Universidad Tecnol\'ogica de Bol\'ivar,\\ Cartagena 13001, Colombia}

	\author{C. S. L\'opez-Monsalvo}
		\email{cesar.slm@correo.nucleares.unam.mx}
		\affiliation{Instituto de Ciencias Nucleares, Universidad Nacional Autonoma de M\'exico,\\
A.P. 70-543, 04510 Mexico D.F., Mexico}

\author{F. Nettel}
		\email{fnettel@ciencias.unam.mx}
		\affiliation{Departamento de F\'isica, Facultad de Ciencias,\\ Universidad Nacional Aut\'onoma de M\'exico, A.P. 50-542, M\'exico D.F. 04510, Mexico}

	\begin{abstract}
We present a class of thermodynamic systems with constant thermodynamic curvature which, within the context of geometric approaches of thermodynamics, can be interpreted as constant thermodynamic interaction among their components. In particular, for systems constrained by the vanishing of the Hessian curvature we write down the systems of partial differential equations. In such a case it is possible to find a subset of solutions lying on a circumference in an abstract space constructed from the first derivatives of the isothermal coordinates. We conjecture that solutions on the characteristic circumference are of physical relevance, separating them from those of pure mathematical interest. We present the case of a one-parameter family of fundamental relations that -- when lying in the circumference --  describe a polytropic fluid. 
	\end{abstract}

\maketitle

\section{Introduction}

The study of physical systems that admit a geometric description in terms of Riemannian manifolds is an interesting and timely subject.  Over the last few years, there has been a number of efforts towards an integrated description of thermodynamics in terms of Legendre invariant quantities. In particular, analogously to the case of field theories, it has been argued that the curvature of the appropriate manifold should be linked to the notion of thermodynamic interaction \cite{quevedo}. There has been numerous proposals in this direction. On the one hand, there are the conformally related metric theories of Ruppeiner and Weinhold , where the metric takes the form of a Hessian of the extensive parameters in the entropy and energy representations, respectively \cite{ruppeiner,weinhold}. However, both fail to comply with the spirit of the geometric construction of field theories, i.e. those are \emph{not} invariant under the natural set of transformations in thermodynamics. On the other hand, the Geometrothermodynamics programme (GTD) has successfully managed to provide us with a set of metrics which are independent of the potential used \cite{quevedo} and the fundamental representation one uses \cite{conformal,applications}.

In the GTD programme one posits that the physical information about a thermodynamic system cannot depend on the potential used to describe it and that such information is encoded in the curvature of the maximal integral manifold of the Pfaffian system defining the first law of thermodynamics [c.f. equation \eqref{theta}]. We call such a manifold the \emph{space of equilibrium states}. The curvature of such manifold is obtained from the first fundamental form, induced but a Legendre invariant metric of the contact manifold specified by the Pfaffian form. In section \ref{sec.gtd} we present a brief review of the programme, in particular, we centre our attention on  a metric whose curvature does not depend on the fundamental representation. 

Thus far, the GTD formalism has been applied to a number of thermodynamic systems in order to test the consistency of the programme. In this work, we take a step forward and set ourselves to the task of finding those systems which exhibit constant \emph{thermodynamic} interaction. That is, we will find the class of fundamental functions producing a manifold of constant curvature. We will restrict ourselves to the two-dimensional case, i.e. we consider only systems with two degrees of freedom. These systems are interesting from the mathematical  point of view since the any two dimensional metric can be cast into a conformally flat form.

The paper is organised as follows. In section II we present a brief account of the Geometrothermodynamics programme and two-dimensional thermodynamic systems of constant curvature. In section III we analyse the system of partial differential equations to find the set of isothermal coordinates for metrics with vanishing Hessian curvature. There we propose a criterion to single out physical fundamental relations based on a circumference-like equation in an abstract space related to the system of differential equations for the isothermal coordinates. To close this section we present an example illustrating these matters. Finally, in section IV we present a summary of the results and address future work on the subject.

\section{Two-dimensional thermodynamic systems of constant curvature}
\label{sec.gtd}

The GTD programme promotes the natural formalism of thermodynamics in terms of contact manifolds to a Legendre invariant Riemannian structure. Let us begin with a brief review of the programme by considering the case of two thermodynamic degrees of freedom. In this case, we need a five dimensional manifold which admits a set of local coordinates  corresponding to the collection of extensive and intensive variables -- denoted  by $q^i$ and $p_i$, respectively -- together with the thermodynamic potential, $\Phi$, such that the kernel of the 1-form
	\beq
	\label{theta}
	\Theta = \d \Phi - p_1 \d q_1 - p_2 \d q_2 ,
	\eeq
generates a maximally non-integrable set of hyperplanes, $\xi \subset T\mathcal{T}$.   A manifold $\mathcal{T}$ together with the 1-form $\Theta$ is called a contact manifold. In the present case we refer to it as the \emph{thermodynamic phase-space}.  Of special interest is the maximal integral sub-manifold, $\mathcal{E} \subset \mathcal{T}$, i.e. the largest sub-manifold which can be embedded in $\mathcal{T}$ such that $T\mathcal{E} \subset \xi$. It is easy to see that this is a two-dimensional manifold  defined by the first law of thermodynamics
	\beq
	\d \Phi = p_1 \d q_1 + p_2 \d q_2, \quad \text{where} \quad \Phi = \Phi(q_1,q_2) \quad \text{and} \quad p_i = \frac{\partial \Phi}{\partial q_i}\equiv \Phi_{,i}.
	\eeq
Thus, we see that if we know the fundamental function $\Phi= \Phi(q_1,q_2)$, then we know how $\mathcal{E}$ is embedded in $\mathcal{T}$. We call the sub-manifold $\mathcal{E}$ the \emph{space of equilibrium states}. 

In addition, the GTD programme introduces a metric structure for the thermodynamic phase space. Such a structure is constructed in order to satisfy the criterion of Legendre invariance, i.e. Legendre transformations correspond to isometries. Within the GTD programme there have been two distinct classes of metrics which have been studied according to their invariance properties, those which are invariant under every possible Legendre transformation and those which are only invariant under \emph{total} Legendre transformations. The metric structure of $\mathcal{T}$ induces a Riemannian metric on $\mathcal{E}$, its first fundamental form, whose intrinsic curvature is associated with the thermodynamic interaction of the system. In our two-dimensional scenario, this whole information is contained in the curvature scalar of $\mathcal{E}$.

If the curvature of the space of equilibrium states is to give a faithful account of the thermodynamic interaction, it should not depend on the choice of fundamental representation, i.e. one is free to work in the energy or entropy representation indistinctly. It has been shown that the metric compatible with both, Legendre and representation invariance is
		\beq
	\label{gnatural}
	G^\natural = \Theta \otimes \Theta + \frac{1}{{q_2} p_{2}} \left( \d q_1 \otimes \d p_1 + \d q_2 \otimes \d p_2 \right).  
	\eeq
Thus, the induced metric on $\mathcal{E}$ is simply given by
	\beq
	\label{gnat}
	g^\natural = \Omega(q_1,q_2) h.
	\eeq
Here $h$ is the Hessian metric 
	\beq
	h = \Phi_{,11} \ \d q_1 \otimes \d q_1  + \left( \Phi_{,12} + \Phi_{,21} \right)\ \d q_1 \otimes \d q_2 + \Phi_{,22}\ \d q_2 \otimes \d q_2,
	\eeq
where we have used a coma to denote partial differentiation with respect to the corresponding coordinate function of $\mathcal{E}$ and the conformal factor is given by 
	\beq
	\Omega(q_1,q_2) = \frac{1}{q_2 \Phi_{,2}}.
	\eeq
The interested reader in the derivation of the metric \eqref{gnat} is referred to \cite{conformal} and to \cite{applications} for applications to ordinary thermodynamic systems.

Note that the components of the metric \eqref{gnat} depend on the second derivatives of the fundamental function $\Phi$ but are otherwise unspecified. It is an interesting exercise to find a class of fundamental functions for which the space of equilibrium states $\mathcal{E}$ becomes an Einstein manifold for the metric \eqref{gnat}. That is, we look for solutions of the system
	\beq
	\label{eins1}
	R^\natural_{\ ab} = K g^\natural_{\ ab},
	\eeq
 where $R^\natural_{\ ab}$ is the Ricci tensor associated with $g^\natural$ and $K$ is a constant, which in the present case corresponds to the Gaussian curvature of $\mathcal{E}$. 
 
It is worth noting that equation \eqref{eins1} represents a system of three, \emph{third} order, non-linear partial differential equations for the thermodynamic potential $\Phi$. Indeed, it is straightforward to show that in two dimensions, the fourth order terms in the curvature exactly cancel whenever the metric is the Hessian of a scalar function.
 
We can reduce the system \eqref{eins1}  by raising one of the indexes to obtain
	\beq
	\label{einseq0}
	{R^\natural}_{a}^{\ b} = K \delta_a^{\ b}.
	\eeq
Thus, the system reduces to the single PDE
	\beq
	\label{einseq}
	F(\Phi_{,i},\Phi_{,ij},\Phi_{,ijj},\Phi_{,iii}) = 4 K \frac{\rho^2}{\Omega^5(q_1,q_2)} \quad \text{with}\quad i,j=1,2.
	\eeq
Here $\rho$ is the determinant of the metric \eqref{gnat} given by the expression
	\beq
	\rho = \Omega^{2}(q_1,q_2) \left(\Phi_{,11}\Phi_{,22} - \Phi_{,12}^2 \right),
	\eeq
andthe lhs of \eqref{einseq} is
	\begin{align}
	\label{eqlarge}
	F(\Phi_{,i},\Phi_{,ij},\Phi_{,ijj},\Phi_{,iii}) = & \ \Phi_{,2}^2 \left(A\ \Phi_{,11}^2 + B\ \Phi_{,11} - 2 q_2 \Phi_{,211}\Phi_{,12}^2 + C\ \Phi_{,12} + D\ \Phi_{,22} \right) \nonumber\\
		& + \Phi_{,2} \left(q_2^2 \Phi_{,22}\Phi_{,222}\Phi_{,11}^2 + E\ \Phi_{,11} + 2 q_2^2 \Phi_{,12}^3 \Phi_{,221} - q_2^2 \Phi_{,12}^2 \Phi_{,22}\Phi_{,211}\right)\nonumber\\
		& - 2 q_2^2\ \rho\ \Omega^{-2} \left( \Phi_{,22}^2 \Phi_{,11} - \Phi_{,12}^2 \Phi_{,22}  \right),
	\end{align}
where 
	\begin{align}
	A = &\ - q_2\Phi_{,222} - 2 \Phi_{,22}, \\
	B = &\ 2\Phi_{,12}^2 + 3 q_2 \Phi_{,12}\Phi_{,221} - q_2 \Phi_{,211} \Phi_{,22} + q_2^2 \Phi_{,221}^2 - q_2^2\Phi_{,222}\Phi_{,211},\\
	C = &\ -q_2^2 \Phi_{,221}\Phi_{,211} + q_2^2 \Phi_{,111}\Phi_{,222} + q_2 \Phi_{,22}\Phi_{,111},\\
	D = &\ -q_2^2 \Phi_{,111}\Phi_{,221} + q_2^2 \Phi_{211}^2,\\
	E = &\ -2 q_2^2 \Phi_{,22}\Phi_{,12}\Phi_{,221} - q_2^2 \Phi_{,12}^2 \Phi_{,222} + q_2^2\Phi_{,22}^2\Phi_{,211}.
	\end{align}
	
Motivated by the results of a previous work by the authors (c.f. Section III.D in \cite{applications} and \cite{aztlan}), we know that a solution to \eqref{einseq0} is given by the fundamental relation 
	\beq
	\label{chap}
	\Phi = \Phi_0 \log\left(q_1^{\alpha} + c\ q_2^{\alpha} \right).
	\eeq
Since we are working in two dimensions, the Gaussian and scalar curvature are proportional and we see that the constancy of $K$ is satisfied  and has the value
	\beq
	\label{constantk}
	K =- \frac{1}{4} \frac{\alpha^2}{\alpha - 1}.
	\eeq

Thus, we can propose a general solution of the form
	\beq
	\label{gen.ansatz}
	\Phi = f(\xi q_1 + \chi q_2).
	\eeq
Here $f$ is a sufficiently differentiable function of the sum of the extensive parameters where $\xi$ and $\chi$ are constants. This type of ansatz does solve \eqref{eqlarge}. However, a quick inspection to the metric determinant reveals the degeneracy of this case [c.f. equation \eqref{metdet}, below]. Therefore, let us propose the more general solution
	\beq
	\label{gensol}
	\Phi = f\left(\xi \ q_1^{\alpha} + \chi \ q_2^{\alpha} \right), 
	\eeq
where $\alpha$ is a constant different from one. Now, the metric determinant is in general different from zero and has the form
	\beq
	\label{metdet}
	\rho = \frac{a-1}{\chi\ q_2^{2+a}\ q_1^2 f'} \left[f' + \xi a q_1^a \left(\xi \ q_1^{\alpha} + \chi\ q_2^{\alpha} \right) f'' \right],
	\eeq
where $f'$ and $f''$ are the first and second total derivatives of the fundamental relation \eqref{gensol} evaluated at $(\xi\ q_1^{\alpha} + \chi\ q_2^{\alpha})$. Now we can clearly see the degeneracy for $\alpha = 1$.

Substituting our ansatz \eqref{gensol} into the equation \eqref{einseq0} we obtain again the same result as in the case of \eqref{chap}, i.e the Gaussian curvature is the same constant, equation \eqref{constantk}, regardless of the particular form of the function $f$ as long as the argument is $\left(\xi \ q_1^{\alpha} + \chi\ q_2^{\alpha} \right)$. Therefore, the generalised Chaplygin gas, equation \eqref{chap} belongs to a class of thermodynamic systems with the same type of interaction given by \eqref{gensol}. Moreover, the Hessian metric for the logarithmic form of this type of fundamental relation has vanishing curvature. In this case it becomes a simpler problem  to find the set of isothermal coordinates for the space of equilibrium states.

\section{Isothermal coordinates}

It is a well known result that every two-dimensional Riemannian manifold is conformally flat. That is, we can always find a set of coordinates for which the metric takes the form
	\beq
	\label{confflat}
	g = \tilde\Omega^2(x,y) g^\flat, \quad \text{where} \quad g^\flat = \d x \otimes \d x + \d y \otimes \d y.
	\eeq
Such a coordinate system is called \emph{isothermal}. In this section we find the isothermal coordinates for the space of equilibrium states $(\mathcal{E},g^\natural)$ under the assumption of the Hessian flatness, i.e. by demanding that the curvature scalar of the Hessian part of the metric \eqref{gnat} vanishes.

Let us consider the diffeomorphism $\varphi:\mathcal{E} \rightarrow \mathcal{E}$  accounting for the change of coordinates $x=x(q_1,q_2)$ and $y = y(q_1,q_2)$. Then we can pull-back the metric $g^\flat$ [c.f. equation \eqref{confflat}] and solve the equation
	\beq
	\label{eq:pb}
	\varphi^*g^\flat - h = 0
	\eeq
for the coordinate functions $x$, $y$ and the thermodynamic potential $\Phi$. This will provide us with a family of thermodynamic fundamental relations with zero Hessian curvature together with their isothermal coordinates. Equation \eqref{eq:pb} above corresponds to the system of equations
 	\begin{align} 
	x_{,1}^2 + y_{,1}^2 & = \Phi_{,11}, \label{I.system} \\
	x_{,1} x_{,2} + y_{,1} y_{,2} & = \Phi_{,12}  \label{II.system}, \\
	x_{,2}^2 + y_{,2}^2 & = \Phi_{,22} \label{III.system}.
 	\end{align}
 	
There is a large family of solutions for such a system. In particular, for separable fundamental relations,
	\beq
	\Phi = S(q_1) + T(q_2), 
	\eeq
 we have that the change of coordinates is given by
	\begin{align}
	x = & \int \sqrt{ \left(1-c^2\right) S''}\ \d q_1 + c \int \sqrt{\ddot T}\ \d q_2,\\
	y = & \int \sqrt{\left(1-c^2\right) \ddot T}\ \d q_2 - c \int \sqrt{ S''}\ \d q_1,
	\end{align}
 where $c$ is a constant and the primes and dots denote differentiation with respect to $q_1$ and $q_2$, respectively.
 
 A more general solution is given by those fundamental functions satisfying the third order system of PDE's
	\beq
	\label{gensoln1}
	\Phi_{,211}  =  \frac{\Phi_{,222} \Phi^2_{,12}}{\Phi_{,22}^2} \quad \text{and} \quad\Phi_{,1,22} = \frac{\Phi_{,222} \Phi_{1,2}}{\Phi_{22}}.
	\eeq
 In this case, the isothermal coordinates  must satisfy the system
	\begin{align}
	\label{eqx1}
	x_{,1}^2 &= \frac{1}{\Phi_{,22}} \left(2 \Phi_{,12} x_{,1}x_{,2} + \Phi_{,11}\Phi_{,22} - x_{,2}^2 \Phi_{,11} - \Phi_{,12}^2 \right),\\
	\label{eqx2}
	x_{,22}	 &= \frac{1}{2} \frac{\Phi_{,222}x_{,2}}{\Phi_{,22}}, \quad \text{and}\\
	\label{yint1}
	y 		 &= \int \sqrt{\Phi_{,22} - x_{,2}^2} \d q_2 + \frac{1}{2}\int \frac{1}{\sqrt{\Phi_{,22} - x_{,2}^2}} \left[\sqrt{\Phi_{,22} - x_{,2}^2}\int \frac{2 x_{,2}x_{,12} - \Phi_{,221}}{\sqrt{\Phi_{,22}- x_{,2}^2}} \d q_2 -2x_{,1} x_{,2} + 2 \Phi_{,12}\right] \d q_1.
	\end{align}

One can verify that a fundamental relation of the form \eqref{gen.ansatz} is a solution of \eqref{gensoln1}. We have seen that this type of functions generates degenerate Hessian metrics. However, we can use them to learn some properties about the space of solutions of the system \eqref{I.system} - \eqref{III.system}. For example, consider the fundamental relations given by
	\beq
	\label{soltest1}
	\Phi = \log(\xi q_1 + \chi q_2).
	\eeq
 In this case we can solve the pair of equations for $x$, i.e. equations \eqref{eqx1} and \eqref{eqx2} to obtain 
	\beq
	 x = c \log\left(q_2 + \frac{\xi}{\chi}q_1 \right),
	\eeq
and substitution in \eqref{yint1} yields
	\beq
	y = \sqrt{- (1 + c^2)} \log(\xi q_1 + \chi q_2).
	\eeq
Thus we see that, indeed, this type of fundamental relation fails to produce a real change of coordinates satisfying \eqref{eq:pb}. Moreover, note that we can find particular solutions to the system \eqref{I.system} - \eqref{III.system} if we restrict ourselves to a circumference in an abstract XY plane. Thus, we have 
    \beq
    \label{cir.eq}
    X^2 + Y^2 = R^2,
    \eeq
where
	\begin{align}
	\label{sys.1}
	X^2 = & (x_{,1} + x_{,2})^2\\
	Y^2 = & (y_{,1} + y_{,2})^2\\
	\label{sys.3}
	R^2 = & \Phi_{,11} + 2 \Phi_{,12} + \Phi_{,22}.
	\end{align}
From this point of view, we observe that the fundamental relation \eqref{soltest1} corresponds to an `imgainary' radius of the circumference \eqref{cir.eq}, that is
	\beq
	\label{radius}
	R^2 = - \frac{(\xi + \chi)^2}{(\xi q_1 + \chi q_2)^2}.
	\eeq
This is not surprising since we knew that the Hessian corresponding to this fundamental relation is degenerate and thus the system is not well posed except for the case $\xi = -\chi = 1$, for which $R=0$.
	
We can use this geometric construction to probe the space of solutions for a \emph{fixed} fundamental relation by noting that a solution to the system \eqref{I.system} - \eqref{III.system}  must lie on the circumference associated with the particular fundamental relation we use [c.f. equation \eqref{radius}], but not every solution lying on the circumference solves the system we are probing.

\subsection{Example}

To see how this construction works,  let us chose a family of fundamental relations  in the form of equation \eqref{chap} parametrised by the exponent $\alpha$. We work in the entropy representation using molar quantities. Thus we set by $q_1 = u$ the specific energy and $q_2 = v$  is the specific volume. The fundamental relation is written as 
	\beq
	s_\alpha = \log\left( u^{\alpha} +  v^{\alpha} \right).
	\eeq
Each of these  functions defines a Hessian metric of zero curvature and a natural metric of constant thermodynamic interaction [c.f. equation \eqref{constantk}]. The change to isothermal coordinates for this type of functions cannot be expressed analytically. However, we can use the circumference to classify the various types of differential equations obtained for each value of $\alpha$.

\begin{figure}
	\begin{center}
	\includegraphics[width=0.49\columnwidth]{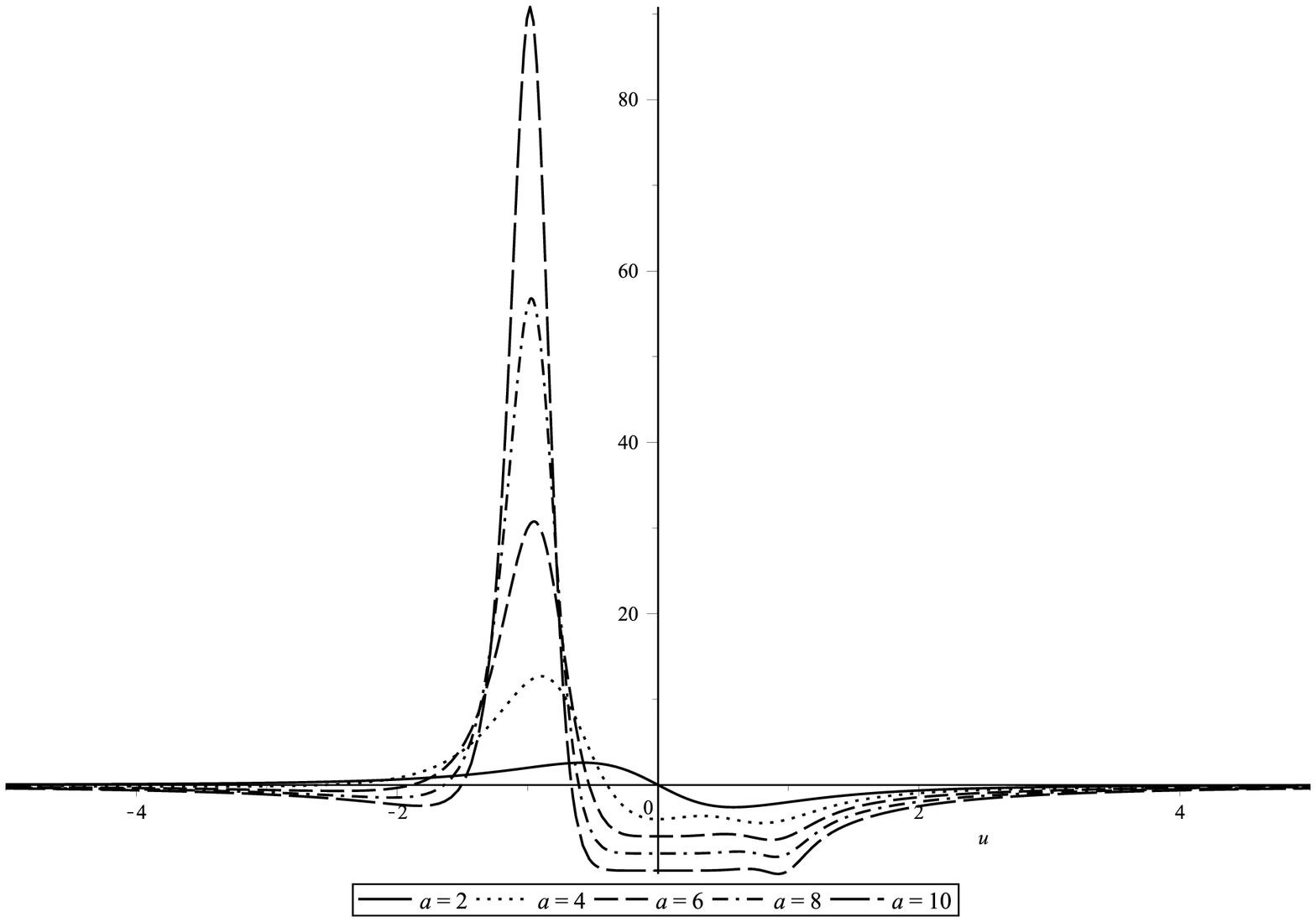}	\includegraphics[width=0.49\columnwidth]{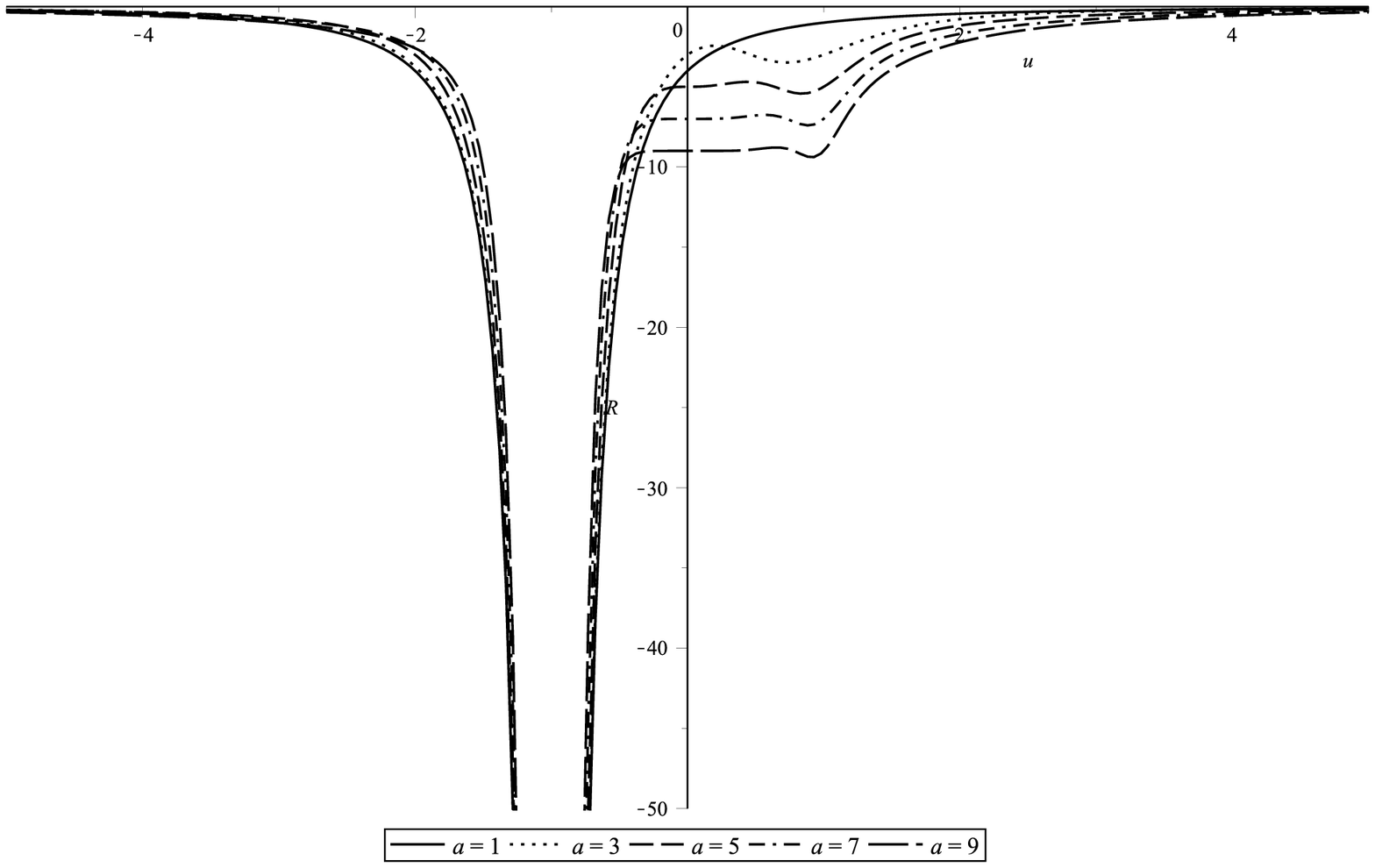}\\
	\includegraphics[width=0.49\columnwidth]{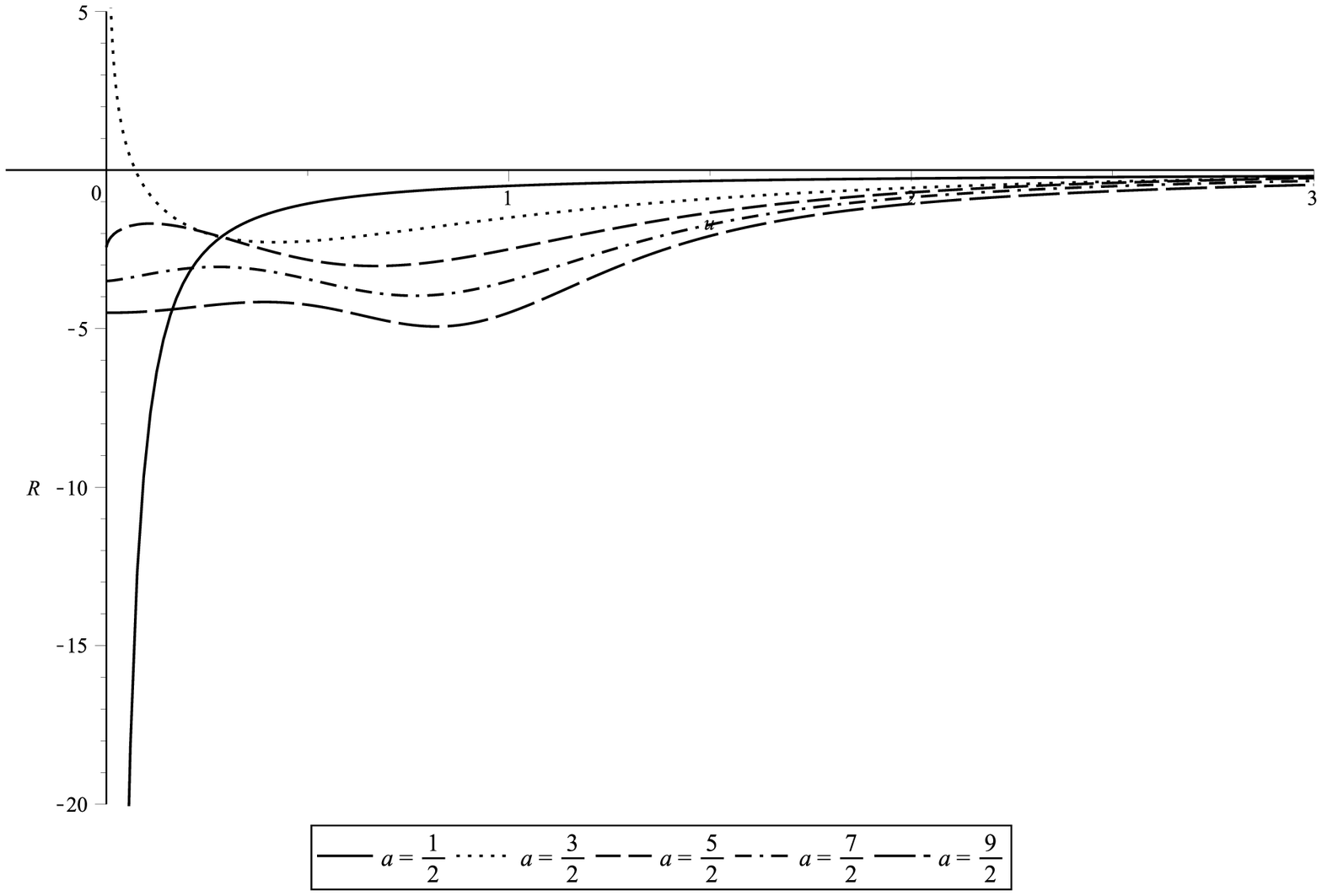}	\includegraphics[width=0.49\columnwidth]{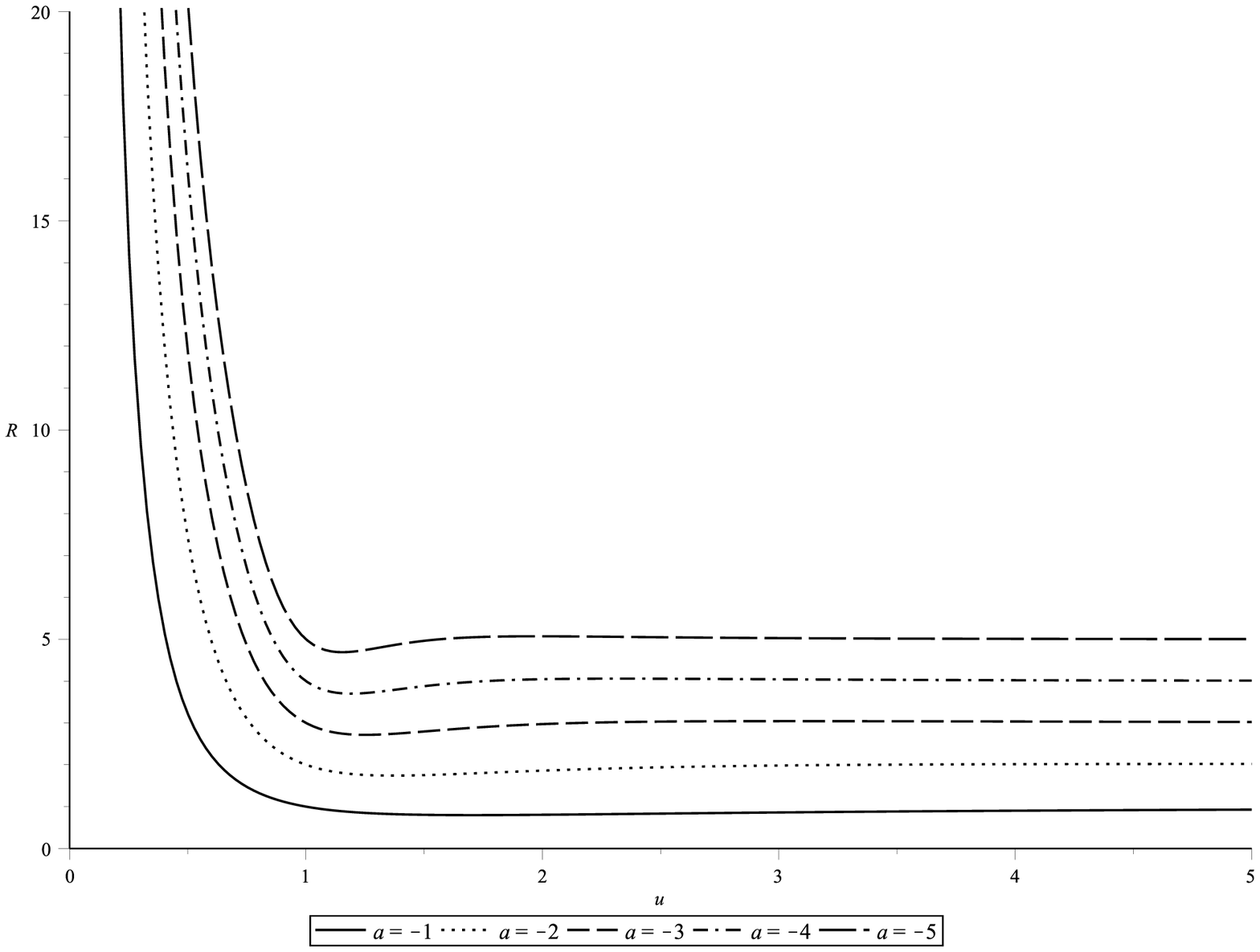}
	\end{center}
	\caption{The three qualitatively different types of radial functions. The horizontal axis corresponds to the energy, while the vertical represents the radius of the circumference. We observe that only the curves corresponding to negative values of the exponent $\alpha$ are positive definite in the full physical domain, i.e. positive values of energy and volume.  }
	\label{fig1}
	\end{figure}

 The squared radii of the circles associated with each function is given by
	\beq   \label{radius1}
	R^2_\alpha = - \frac{\alpha}{u^2 v^2 \left( u^\alpha +  v^\alpha\right)} \left[ u^{2\alpha} v^2 -  v^\alpha u^\alpha \left((\alpha-1) u^2 - 2 \alpha u v + v^2 (\alpha -1) \right) +  v^{2\alpha}u^2\right]
	\eeq
Thus, fixing the volume of the system, we find the subset of fundamental functions for which the PDE represented by the circumference \eqref{cir.eq} is well defined. The three qualitatively different types of behaviour are depicted in figure \ref{fig1}. It is easy to observe that for $\alpha >0$ the PDE is ill defined, thus, only fundamental relations for which $\alpha < 0$ are meaningful. This corresponds to a positive thermodynamic curvature as can be seen from \eqref{constantk}. Furthermore, noticing the symmetry of $u$ and $v$ in the expression for the circumference radius \eqref{radius1}, it is easy to observe that the PDE also restrict the domain of the thermodynamic variables to $u, v > 0$. The fundamental relation with $\alpha < 0$ describes a polytropic fluid with equation of state given by
	\beq  \label{eos}
	P = \rho^{1-\alpha},
	\eeq
where $\rho = \frac{u}{v}$ is the energy density of the fluid.  It is a simple task to obtain the heat capacity at constant volume for these systems
	\beq \label{cv}	
	c_v = \frac{a u^\alpha}{u^\alpha + (1-\alpha) v^\alpha}.
	\eeq
From this expression we observe that the heat capacity remains finite for any value of the thermodynamic variables and is always negative whenever $\alpha < 0$.

		\begin{figure}
	\begin{center}
	\includegraphics[width=0.49\columnwidth]{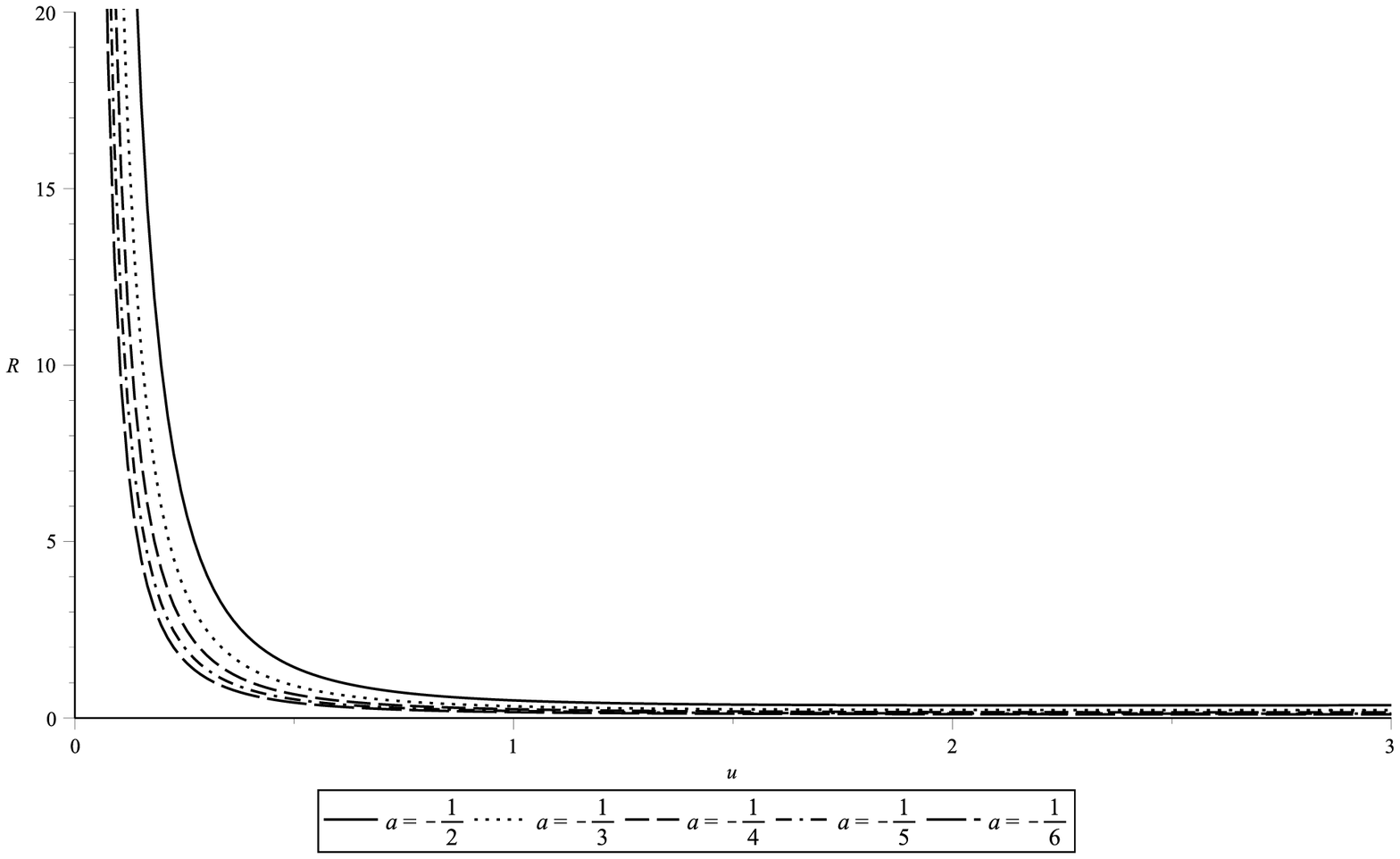}	\includegraphics[width=0.49\columnwidth]{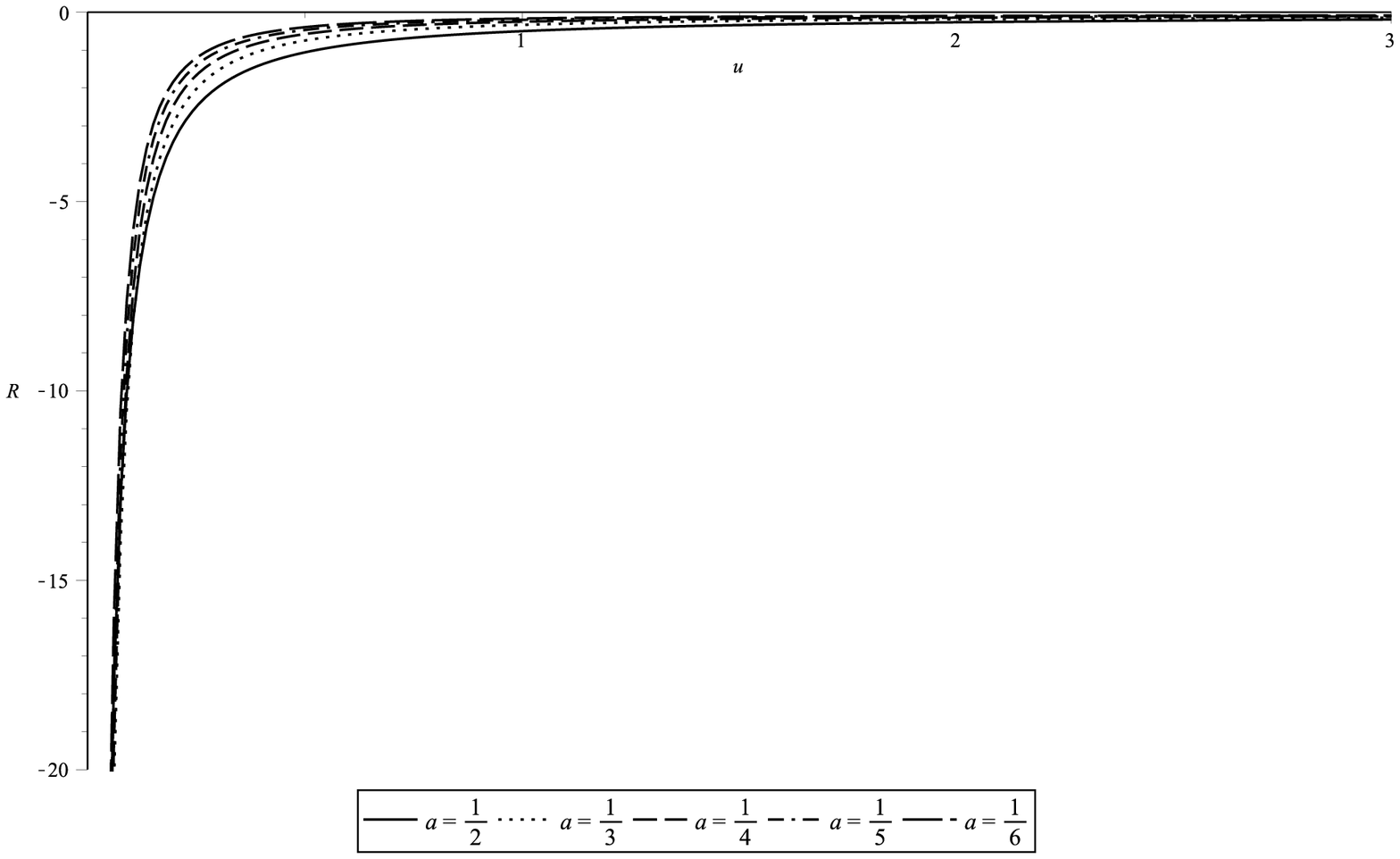}\\
	\end{center}
	\caption{These plots correspond to the radial function for $-1\leq \alpha \leq 1$. We observe that only those values of $\alpha<0$ can be associated with physical systems. }
	\label{fig2}
	\end{figure}

\section{Closing remarks}

In this paper we studied two-dimensional Einstein manifolds for the Geometrothermodynamics programme. We found the differential equation that must be satisfied by the fundamental relation in order to describe a system with constant thermodynamic interacion, i.e. a fundamental relation producing a representation invariant metric whose associated curvature is constant. In particular, building on previous work (c.f. reference \cite{applications}) we analysed the one-parameter family of fundamental relations given by equation \eqref{gensol}. 

Noting the conformal structure of $g^\natural$ [see equation \eqref{gnat}], we centre our study in the class of functions whose associated Hessian metric has vanishing curvature. With this assumption, we set up the system of differential equations defining the set of isothermal coordinates. As expected, we found an analytic expression for the change of coordinates for the case in which the fundamental function is separable. An interesting exercise allowed us to explore some properties of the space of solutions of \eqref{eq:pb}, i.e. the set of isothermal coordinates  with their corresponding  fundamental relation of vanishing Hessian curvature.

Observing the algebraic structure of the system \eqref{I.system} - \eqref{III.system}, we note that there will be a class of fundamental relations satisfying the Hessian curvature constraint  for which  we can build a characteristic circumference on which the solutions of the PDE system lie.  This can be done whenever the derivatives of the fundamental relation define a positive squared radial function. Moreover, we conjecture that only the set of fundamental relations for which such a construction is possible can describe physical systems of constant thermodynamic interaction. In particular, we work out the example given by \eqref{chap}. Here, we work in specific thermodynamic variables in the entropy representation of a system characterised by the exponent $\alpha$. Indeed, only those systems for which $\alpha < 0$ correspond to a polytropic fundamental relation.

In sum, we have analysed a particular class of fundamental relations of constant thermodynamic curvature. It remains to explore the larger class of functions within the set of  solutions of equation \eqref{eq:pb} and study their thermodynamic implications. This will be done in a forthcoming article.

\section*{Acknowledgements}

ACGP is funded by a TWAS-CONACYT postdoctoral grant. CSLM is thankful to CONACYT, postdoctoral Grant No. 290679\_UNAM. FN acknowledges support from DGAPA-UNAM (postdoctoral fellowship)


\begin{thebibliography}{99}




\bibitem{quevedo}
H. Quevedo,  
{\it   Geometrothermodynamics}, J. Math. Phys. {\bf 48}, 13506 (2007).


\bibitem{ruppeiner}
G. Ruppeiner,
{\it Thermodynamics: A Riemannian geometric model}, Phys. Rev. A {\bf 20}, 1608 (1979).

\bibitem{weinhold}
F. Weinhold,
{\it Metric geometry of equilibrium thermodynamics}, J. Chem. Phys. {\bf 63}, 2479 (1975).



\bibitem{conformal}
A. Bravetti, C. S. Lopez-Monsalvo, F. Nettel and H. Quevedo,
{\it The conformal metric structure of geometrothermodynamics}, J. Math. Phys. {\bf 54}, In press (2013); arXiv:1302.6928 [math-ph].

\bibitem{applications}
H. Quevedo, F. Nettel, C. S. Lopez-Monsalvo and A. Bravetti,
{\it Representation invariant Geometrothermodynamics: applications to ordinary thermodynamic systems},  (2013); arXiv:1303.1428 [math-ph].




\bibitem{aztlan}
A. Aviles, A. Basterrechea-Almodovar, L. Campuzano and H. Quevedo,
{\it Extending the generalized Chaplygin gas model by using geometrothermodynamics}, Phys. Rev. D {\bf 86}, 063508 (2012).







\end{thebibliography}
\end{document}